\documentclass[oldversion]{aa}  
\usepackage{natbib}
\usepackage{graphicx}

\usepackage{txfonts}
\begin{document}

\titlerunning{Small-scale dynamo action during the formation of the first stars and galaxies.  I. The ideal MHD limit.}
\authorrunning{Schleicher et al.}
   \title{Small-scale dynamo action during the formation of the first stars and galaxies. I. The ideal MHD limit.}


   \author{Dominik R. G. Schleicher
          \inst{1,2}
          \and
          Robi Banerjee\inst{3}
          \and
          Sharanya Sur\inst{3}
          \and
          Tigran~G.~Arshakian\inst{4}
           \and\\
           Ralf S. Klessen\inst{3,5}
           \and
           Rainer Beck\inst{4}
           \and
          Marco Spaans \inst{6}
                    }

   \institute{              ESO, Karl-Schwarzschild-Strasse 2, 85748 Garching bei M\"unchen, Germany;
              \email{dschleic@eso.org}
              \and
              Leiden Observatory, Leiden University, P.O. Box 9513, NL-2300 RA Leiden, the Netherlands
         \and
         Zentrum f\"ur Astronomie der Universit\"at Heidelberg, Institut f\"ur Theoretische Astrophysik, Albert-Ueberle-Str.~2, 69120 Heidelberg, Germany
\and
Max-Planck-Institut f\"ur Radioastronomie, Auf dem H\"ugel 69, 53121 Bonn, Germany
\and
Kavli Institute for Particle Astrophysics and Cosmology, Stanford University, Menlo Park, CA 94025, U.S.A.
\and
             Kapteyn Astronomical Institute, University of Groningen, P.O. Box 800, 9700 AV, Groningen, the Netherlands
             }

   \date{Received September 15, 1996; accepted March 16, 1997}

\abstract{We explore the amplification of magnetic seeds during the formation of the first stars and galaxies. During gravitational collapse, turbulence is created from accretion shocks, which may act to amplify weak magnetic fields in the protostellar cloud. Numerical simulations showed that such turbulence is sub-sonic in the first star-forming minihalos, and highly supersonic in the first galaxies with virial temperatures larger than $10^4$~K. We investigate the magnetic field amplification during the collapse { both for Kolmogorov and Burgers-type} turbulence with a semi-analytic model that incorporates the effects of gravitational compression and small-scale dynamo amplification. We find that the magnetic field may be substantially amplified { before the formation of a disk. On scales of $1/10$ of the Jeans length, saturation occurs after $\sim10^8$~yr. } Although the saturation behaviour of the small-scale dynamo is still somewhat uncertain, we expect a saturation field strength of the order  $\sim 10^{-7} n^{0.5}$~G in the first star-forming halos, with $n$ the number density in cgs units. In the first galaxies with higher turbulent velocities, the magnetic field strength may be increased by an order of magnitude{, and saturation may occur after $10^6-10^7$~yr. In the Kolmogorov case, the magnetic field strength on the integral scale (i.e. the scale with most magnetic power) is higher due to the characteristic power-law indices, but the difference is less than a factor of $2$ in the saturated phase. Our results thus indicate that the precise scaling of the turbulent velocity with length scale is of minor importance. They further imply that magnetic fields will be significantly enhanced before the formation of a protostellar disk, where they may change the fragmentation properties of the gas and the accretion rate.}
}

               
   \maketitle
%

\section{Introduction}
The formation of the first stars is generally regarded as a well-defined problem, as the initial conditions at $z\sim100$ can be derived accurately
from CMB data \citep[e.g.][]{Komatsu09} using linear theory \citep{Bertschinger98} and the chemistry is primordial and well-understood \citep[e.g.][]{Abel97,Galli98, Stancil98, Omukai01, Yoshida06, Schleicher08, Glover09}. In addition, it is often assumed that magnetic fields are not yet present 
and that the hydrodynamical equations are sufficient to describe the star formation process. 

This assumption is not neccessarily true. Indeed, a variety of mechanisms exist to create strong magnetic fields during inflation, the electroweak or the QCD phase transition \citep[see e.g.][for a review]{Grasso01}. By means of the inverse-cascade, the magnetic power of these fields may have been shifted to larger scales in case of non-zero helicity \citep{Brandenburg96, Christensson01,Banerjee04b}. Strong primordial fields  would have profound implications concerning the thermodynamics of the post-recombination universe, reionization and the formation of the first stars \citep{Sethi05, Machida06, TashiroSugiyama06a, Tashiro06b, Schleicher08b, Schleicher09a, Schleicher09prim}. 

In this paper, we will however explore the limiting case in which extremely weak seed fields have been produced before recombination. In such a case, the dominant contribution to the magnetic field strength comes from astrophysical processes after recombination.  Cosmological MHD simulations including an approximate treatment of the Biermann battery term suggest that the Biermann battery could create seed field of the order $10^{-18}$~G in the IGM at $z=20$ \citep{Xu08}. Additional seed fields may be created via the Weibel instability in shocks \citep{Schlickeiser03, Medvedev04, Lazar09}. The importance of dynamos in cosmic sheets has early been recognized by \citet{Pudritz89}.  The simulations of \citet{Xu08} run from cosmological scales to the protostellar collapse phase in a primordial minihalo. In such a situation, the following mechanisms are available to amplify the magnetic field:

\begin{itemize}
\item gravitational compression of the magnetic field,
\item { the small-scale turbulent dynamo which amplifies seed magnetic fields 
already generated from cosmological processes},
\item {  large-scale dynamos in protostellar and galactic disks},
\item the magneto-rotational instability (MRI).
\end{itemize}
Gravitational compression under spherical symmetry leads to an increase of the magnetic field strength with $n^{2/3}$, where $n$ denotes the number density of the gas. If the collapse proceeds preferentially along one axis, for instance because of rotation or strong magnetic fields, the scaling is closer to $n^{0.5}$. In realistic cases, often intermediate values are found \citep{Machida06, Banerjee08}. This amplification mechanism has also been identified in the simulation of \citet{Xu08}.

Large-scale dynamos typically require the presence of a galactic or protostellar disk and act on relatively long timescales \citep[see][for a review]{Brandenburg05}. { In such a disk, an exponential growth may also be obtained from the magnetorotational instability \citep[MRI, see][]{Balbus91},  which may seed other large-scale dynamos with turbulence \citep{Tan04, Silk06}. However, the length scale of the fastest growing mode decreases for decreasing field strengths, and in the presence of a viscous cutoff length, such amplification may not be possible. As a result, a minimal field strength is required to drive the MRI in a protostellar disk \citep[see][for a detailed discussion]{Tan04, Silk06}.}


A critical condition for any dynamo growth is that the ideal MHD approximation is applicable. \citet{Maki04, Maki07} investigated this question using detailed models for magnetic energy dissipation via Ohmic and ambipolar diffusion to show that the magnetic field is frozen into the gas unless it is very strong. An approximate fit to their results yields a critical field strength of about $B\leq10^{-5}(n/10^3\ \mathrm{cm}^{-3})^{0.55}$~G. Due to the subtle effects of lithium chemistry, the ionization degree does not drop exponentially at densities of $\sim10^9$~cm$^{-3}$, but stays almost constant with increasing density. The more recent study by \citet{Glover09} finds even higher ionization degrees at these densities. This implies that the ideal MHD approximation can be used during the collapse phase to describe the interaction of magnetic fields with matter. 

Deviations from this behaviour may however occur on very small scales, where ambipolar and Ohmic diffusion become increasingly important. Estimates based on the non-ideal MHD models of \citet{Pinto08a, Pinto08b} imply that, even if the magnetic field is in equipartition with the gas, ambipolar diffusion is important only on scales 4 orders of magnitudes smaller than the Jeans length, and Ohmic dissipation occurs only on even smaller scales. As Ohmic and ambipolar diffusion depend on the field strength itself, these scales will be significantly smaller for weaker magnetic fields, so that the ideal MHD approximation can be savely applied. This implies that the magnetic Reynolds number ${\rm Rm}=vl/\eta$ varies strongly during the growth of the magnetic field, but always fulfills the condition Rm$\ggg1$ and Pr$_M=\nu/\eta>1$. Detailed calculations concerning the ambipolar and Ohmic diffusion scales will be presented in a companion paper, in which we make use of the ionization degree obtained from a numerical simulation to provide an updated calculation of these scales for different field strengths. 

Gravitational collapse is generally accompanied by the presence of turbulence \citep{Klessen09}, which may for instance be described by the theory of \citet{Kolmogorov41}. Numerical simulations show that primordial star formation during the collapse phase occurs in a self-similar fashion, where the density profile at a given time is always well-described by a Bonner-Ebert sphere with a flat central density core \citep{Abel02, Bromm03, Yoshida08}. Similar results have been found for present-day star formation \citep[e.g.][]{Banerjee06b, Banerjee07a}. The gas falling on these central cores leads to weak shocks up to Mach~$1$, which drive turbulence in the central density core. This is reflected in the inhomogeneities in the central core and the sub-Keplerian angular momentum profiles, as reported by \citet{Abel02, Bromm03, Yoshida08}. Under such conditions, a strong tangled magnetic field may be generated already during the collapse phase by the small-scale dynamo that was originally proposed by \citet{Kazantsev68}.  

This dynamo provides a very generic means of amplifying magnetic fields and was also proposed to be important in the large-scale structure of the universe \citep{Ryu08}. {The field amplification is due to the random stretching and folding of the magnetic field lines in a turbulent random flow. In the kinematic regime, the field grows typically on the eddy turnover 
time, $t_{\rm ed} = l/v$ where $l$ is a typical turbulent length scale and $v$ is the turbulent 
velocity.} { In the context of galaxy formation, \citet{Beck94} proposed that it is the small-scale dynamo that produces the seeds for galactic large-scale dynamos. As pointed out by \citet{Arshakian09}, the small-scale dynamo can effectively amplify weak seed magnetic fields by $\sim {13}$ orders of magnitude on a timescale $\sim 300$ million years in the first galaxies. Similar results were obtained by \citet{deSouza10} from a direct solution of Kazantsev's equation.} 

Capturing such dynamos in numerical simulations of protostellar collapse is extremely challenging, as 
it requires that the turbulent cascade is well-resolved and well-separated from the scale where MHD turbulence is numerically dissipated. State of the art numerical simulations of { turbulence thus require a spatial resolution of at least $512^3$ for a marginally resolved inertial range \citep{Federrath08, Federrath10}.} Numerical simulations following protostellar collapse, on the other hand, typically resolve the Jeans length and thus the high density region with about $16$ cells, rendering them unable to capture the potential amplification via the turbulent dynamo. { \citet{Federrath10} showed that at least 30 grid cells are required to resolve turbulent vortices.} Simulations as performed by \citet{Xu08} therefore cannot  resolve the turbulence in the central core. 

In this paper, we explore the implications of the small-scale dynamo during the gravitational collapse phase within a semi-analytic framework, applied to the formation of the first stars and galaxies. We first review the theoretical background and numerical evidence for the small-scale dynamo in \S~\ref{dynamo}, and present a set of analytic estimates. In \S~\ref{collapse}, we develop a quantitative model concerning the small-scale dynamo action during the collapse process. This model is applied in \S~\ref{application} both to minihalos and atomic cooling halos, taking into account the amount of turbulence that was found in numerical simulations. Phenomenological consequences from the generation of such magnetic fields are discussed in \S~\ref{conclusions}. { In a companion paper \citep{Sur10}, we present numerical simulations confirming the importance of dynamo amplification during gravitational collapse.}

\section{The small-scale dynamo}\label{dynamo}
In this section, we introduce the small-scale dynamo by sketching its analytic derivation as well as numerical simulations that examined its efficiency, to make these results accessible to a broader community. This will be combined with a summary on the most important results and a first estimates concerning its importance in the first galaxies in \S~\ref{estimates}. Readers that are already familiar with dynamo theory or only interested in the most important results may thus directly proceed from \S~\ref{estimates}.

\subsection{Analytical arguments}\label{analytic}
The small-scale dynamo was introduced by \citet{Kazantsev68} in an analytic framework, which was improved subsequently in various works \citep[e.g.][]{Subramanian98, Subramanian99,Brandenburg05,Arshakian09}. For a detailed review we particularly recommend \citet{Brandenburg05}, as we can sketch here only the main steps in deriving the main properties of this dynamo. The induction equation of the magnetic field $\vec{B}$ is given as
\begin{equation}
\frac{\partial\vec{B}}{\partial t}= \nabla \times \left ( \vec{v}\times \vec{B}-\eta \nabla \times \vec{B}\right)\label{induction}
\end{equation}
where $\vec{v}$ is the velocity field of the fluid and $\eta$ the Ohmic or ambipolar resistivity. The velocity field is typically decomposed into a stochastic field $\vec{v}_T$ and a drift component $\vec{v}_D$. The latter describes the change of the fluid velocity due to the Lorentz force. Such a decomposition is reasonable as long as the kinetic energy dominates over the magnetic energy. As we will see, this is always the case in the regime where the small-scale dynamo is operational. In principle, the drift velocity depends on the complex history of field lines that a certain fluid element has seen during its history. To make the problem analytically tractable, however, it is generally assumed that most of the history averages out due to the tangledness of the magnetic field. Then, the drift velocity points at least approximately into the direction of the instantaneous Lorentz force. It is generally approximated as \citep{Kazantsev68,Subramanian98,Brandenburg05}
\begin{equation}
\vec{v}_D=\frac{\tau}{4\pi \rho}\left[ \left(\nabla\times\vec{B}\right)\times\vec{B} \right],
\end{equation}
where $\tau$ is the typical response time of the magnetic field and $\rho$ the gas density. The stochastic field component $\vec{v}_T$, on the other hand, is assumed to be an isotropic, homogeneous, Gaussian random velocity field with zero mean. We further adopt the Markovian approximation, assuming that its correlation function is given as \citep{Kazantsev68,Subramanian98,Brandenburg05}
\begin{equation}
\langle v_T^i(\vec{x},t) v_T^j(\vec{y},s)\rangle=T^{ij}(r)\delta(t-s),
\end{equation}
with $r=|\vec{x}-\vec{y}|$ and $\delta$ denoting the delta distribution function.  The matrix $T^{ij}$ can be decomposed in a longitudinal part $T_L(r)$,
a transverse part $T_N(r)$ and a helical part $C(r)$. In the presence of homogeneous, isotropic and Gaussian turbulence, a weak seed magnetic field will be dragged with the fluid and obey the same properties. We denote the equal-time two-point correlation function of the magnetic field as $M^{ij}(r,t)=\langle B^i(\vec{x},t)B^j(\vec{y},t)$ and decompose it into a longitudinal component $M_L(r,t)$, a transverse component $M_N(r,t)$ and a helical component $H(r,t)$. Defining $r^i=x^i-y^i$ and introducing the Kronecker delta $\delta^{ij}$ and the totally antisymmetric tensor $\epsilon_{ijk}$, one can write \citep{Subramanian98, Brandenburg05}
\begin{equation}
M^{ij}=M_N\left[ \delta^{ij}-\frac{r^ir^j}{r^2} \right]+M_L \frac{r^ir^j}{r^2}+H\epsilon_{ijk}r^k.
\end{equation}
As the magnetic field is divergence free, one can then show that
\begin{equation}
M_N=\frac{1}{2r}\frac{\partial r^2 M_L}{\partial r}.
\end{equation}
From the induction equation (\ref{induction}), one can then derive the following evolution equation for $M_L$ and $H$ \citep{Kazantsev68,Subramanian98, Brandenburg05}:
\begin{eqnarray}
\frac{\partial M_L}{\partial t}&=&\frac{2}{r^4}\frac{\partial}{\partial r}\left(r^4\kappa_N \frac{\partial M_L}{\partial r}\right)+GM_L-4\alpha_N H,\label{evolution}\\
\frac{\partial H}{\partial t}&=&\frac{1}{r^4}\frac{\partial}{\partial r}\left(r^4\frac{\partial}{\partial r}(2\kappa_N H+\alpha_N M_L)\right).
\end{eqnarray}
These equations include a diffusion term with coefficient $\kappa_N=\eta+T_L(0)-T_L(r)+2aM_L(0,t)$. The coefficient $\alpha_N=2C(0)-2C(r)-4aH(0,t)$ describes the strength of the $\alpha$ effect, and the term involving $G=-4((T_N/r)'+(rT_L)'/r^2)$ describes the rapid generation of magnetic fluctuations by velocity shear. The prime denotes here the derivative with respect to $r$.

We will now focus on the case of nonhelical turbulence, with $C(r)=0$ and $H(r,t)=0$, as it is more straightforward to be treated analytically. The formalism can however be applied to helical turbulence. In that case, the turbulent dynamo acts in the same way on small-scales, but creates additional large-scale correlations that can act as seed fields for a large-scale dynamo \citep{Subramanian99}. We are looking for eigenmode solutions of Eq.~(\ref{evolution}) and thus make the ansatz $\Psi(r)\mathrm{exp}(2\Gamma t)=r^2\sqrt{\kappa_N}M_L$. Insertion in Eq.~\ref{evolution} leads to a time independent equation which formally resembles the Schr\"odinger equation of quantum mechanics \citep{Kazantsev68,Subramanian98, Brandenburg05}:
\begin{equation}
-\Gamma \Psi = - \kappa_N \frac{d^2\Psi}{dr^2}+U_0(r)\Psi,\label{Schrodinger}
\end{equation}
Under the assumption that the velocity field is locally divergence-free, one can show that $U_0(r)=T_L''+(2T_L'/r)+\kappa_N''/2-(\kappa_N')^2/(4\kappa_N)+2\kappa_N/r^2$. Of course, the assumption of a divergence-free velocity field is only approximately true for compressible gas. However, a turbulent velocity field can always be decomposed into a divergence-free and a divergent component, and the divergent component is just neglected to simplify the analytic treatment. As discussed in the next subsection, numerical simulations that generally include both components still find the same results concerning the magnetic field amplification timescale.

As a boundary condition, we adopt $\Psi\rightarrow0$ for $r\rightarrow0,\infty$. The solution of Eq.~(\ref{Schrodinger}) depends on the stability of flow of the magnetized fluid, which is described by the magnetic Reynolds number ${\rm Rm}=vl/\eta$, which describes the relative importance of the interaction of the magnetic field with the velocity field compared to Ohmic or ambipolar dissipation. As discussed in the introduction, high magnetic Reynolds numbers can be expected during primordial collapse. For the solutions of Eq.~\ref{Schrodinger}, a critical magnetic Reynolds number ${\rm Rm}_{\rm cr}\sim60$ exists for which 
the solution corresponds to a bound state with $\Gamma=0$ \citep{Kazantsev68, Subramanian98, Subramanian99}.  For ${\rm Rm}>{\rm Rm}_{\rm cr}$, $\Gamma>0$ modes can be excited, leading to an exponential growth of the magnetic field on the eddy turnover timescale $l/v$. Such a magnetic field is then curved on the turbulent length scale $l$. The thickness of the flux ropes is more uncertain and may depend on the critical Reynolds number for the dynamo \citep{Subramanian98} or the resistive scales \citep{Schekochihin04}. 

To obtain the value at which the magnetic field saturates, the evolution equation (\ref{evolution}) must be solved for a stationary state. The analysis shows that the magnetic energy then corresponds to ${\rm Rm}_{\rm cr}^{-1}$ of the kinetic energy. The maximum magnetic field strength that can be obtained in this way is thus given as \citep{Subramanian98}
\begin{equation}
B_{max}=\sqrt{4\pi \rho v^2} {\rm Rm}_{\rm cr}^{-1/2}.\label{maxfield}
\end{equation}
We note that the saturation field strength is still somewhat uncertain; our main intention here is to show that saturation can be reached. We checked that this conclusion does not depend on the precise fraction for the saturation level adopted here.

\subsection{Confirmation from numerical simulations}
As pointed out above, the complexity of modeling magnetic fields, gas dynamics and their mutual interplay forces one to make simplifying assumptions in an analytic treatment. Thus, numerical simulations are required to test the analytic results.
{ Evidence of small-scale dynamo action has so far been confirmed in numerical 
simulations of forced MHD turbulence \citep{Haugen04a, Schekochihin04} both for high as well
as low magnetic Prandtl numbers. Such simulations were able to follow the evolution of the magnetic field even in the nonlinear regime when the Lorentz forces become strong enough to saturate the dynamo. In the simulations of \citet{Haugen04a}, exponential growth of the magnetic field commences once the magnetic Reynolds number exceeds a certain critical value which for
${\rm Pr}_m=1$ is  ${\rm Rm}\sim 35$. This critical value is somewhat larger by a factor 3 compared
to the ${\rm Rm}_{cr}$ obtained from the Kazantsev model, which assumes a delta-correlated 
velocity field. Simulations starting with an initially coherent field find that the magnetic power spectrum follows a $k^{3/2}$ slope
before saturation sets in. The growth rate of the field increases with 
$({\rm Rm}/{\rm Rm}_{cr})^{1/2}$ for higher magnetic Reynolds number. Apart from this, the 
salient feature of the small-scale dynamo is the presence of highly intermittent and structured 
fields. This property has been discussed in detail by \citet{Schekochihin04, Brandenburg05}. The 
magnetic fields appear to be in folds whose length is comparable to the box size and which 
reverses direction at the resistive scale. The growing magnetic field appears highly 
intermittent in the sense that it has strong positive(negative) values  only in a few places in the simulation domain. These results were confirmed at higher resolution \citep{Haugen04b} and for different Mach numbers \citep{Haugen04c}, making a robust case for the efficiency of the small-scale dynamo. 

\subsection{Order-of-magnitude estimates}\label{estimates}

\begin{figure}
\includegraphics[scale=0.5]{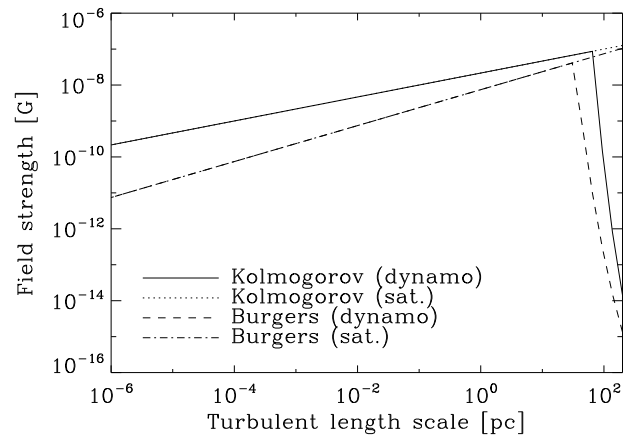}
\caption{Estimates on the field strength that can be reached within a free-fall time as a function of scale. Shown are both the actual field strength to which the magnetic field is amplified by the small-scale dynamo, as well as the maximum field strength at which dynamo action saturates. This saturation field strength increases continuously as a power-law until the scale where turbulence is injected, whereas the actual field strength has a maximum between $10$ and $100$~pc. We consider both Kolmogorov and Burgers turbulence.}
\label{fig:spectrum}
\end{figure}

In summary, previous investigations found that the small-scale dynamo
\begin{itemize}
\item amplifies magnetic fields on the eddy timescale,
\item leads to a magnetic fields curved on scales up to the turbulent length scale,
\item and a thickness of the flux ropes either given from the resistive scales or the critical Reynolds number for dynamo amplification.
\end{itemize}

As the analytic procedures discussed in \S~\ref{analytic} are based on the simplifying assumption that the gas is incompressible, they cannot be directly applied to a situation in which the gas is collapsing. Rather than this, we will focus now on the main results that were obtained with analytical and numerical simulations and develop a new framework where we can apply them during the star formation process.

In order to assess the potential importance of the small-scale dynamo in the first galaxies, we start with an order-of-magnitude estimate concerning the rapid build-up of magnetic fields. The amount of turbulence in the first galaxies has been studied with numerical simulations by \citet{Greif08} and \citet{Wise07a, Wise08a}. On spatial scales of $\sim200$~pc, they find typical turbulent velocities of $\sim20$~km/s. To assess how much such turbulence can amplify magnetic fields in the first galaxies, we need to calculate the number of e-foldings within a free-fall time that are available for magnetic field amplification. The free-fall timescale is given as $t_{ff}=1/\sqrt{\rho G}$, with $\rho$ the total mass density and $G$ Newton's constant. We assume that the turbulent velocity scales as $v(l)\propto l^\beta$, with $\beta=1/3$ for Kolmogorov turbulence and $\beta=1/2$ for Burgers turbulence, on scales smaller than the injection scale. Kolmogorov turbulence describes a situation where the gas is incompressible and is thus applicable in particular for sub-sonic turbulence, whereas Burgers turbulence describes turbulence in the presence of supersonic shocks, where the gas is quite strongly compressed. 

{ We note that Burgers turbulence should be considered as a highly idealized situation, because numerical simulations show that turbulence even in the supersonic regime always consists of rotational and compressional components of comparable strength \citep{Haugen04c, Federrath08, Federrath10}, and dedicated studies by \citet{Kritsuk07}, \citet{Schmidt09} and \citet{Federrath10} typically find power-laws in between the Burgers and Kolmogorov case. As a side note, we mention that purely irrotational turbulence may not be able to drive the small-scale dynamo at all \citep{Mee06}, whereas realistic turbulence should always have rotational and irrotational components as discussed above. 
}


The eddy timescale is given as $t_{\mathrm{ed}}=l/v$. Considering that the small-scale dynamo should not be amplified above the saturation value given in Eq.~(\ref{maxfield}), the magnetic field strength after a free-fall time is given as
\begin{equation}
B = \mathrm{min}\left(B_0\ \mathrm{exp}\left(\frac{v }{l \sqrt{\rho G}}\right), B_{\mathrm{max}}\right)\label{estimate},
\end{equation}
with $B_0$ denoting the strenth of the seed field.  To estimate the potential impact of the small-scale dynamo, we evaluate Eq.~(\ref{estimate}) for a baryonic number density of $0.02$~cm$^{-3}$, { corresponding to the mean density at virialization. We take into account the contribution of dark matter to the total mass density.} According to  \citet{Greif08} and \citet{Wise07a, Wise08a}, we assume that the turbulence is injected on scales of $\sim200$~pc with a velocity of $\sim20$~km/s. On smaller scales, we expect that the turbulent velocity follows some typical scaling law as in Kolmogorov or Burgers turbulence. We explore these cases for definiteness. We adopt a seed field of $B_0\sim10^{-20}$~G, somewhat below the value expected from a Biermann battery mechanism \citep{Xu08}. The results are shown in Fig.~\ref{fig:spectrum}. In case of Kolmogorov turbulence, the magnetic field is negligible on scales of a few hundred parsec, but increases rapidly towards smaller scales. A maximum is reached on scales of { $\sim80$~pc, where the field strength has increased by about $13$ orders of magnitude.} The maximum occurs because the magnetic field has reached the saturation level on this scale. On all smaller scales, the eddy timescale is smaller, decreasing as $l/v\propto l^{2/3}$ in the Kolmogorov case. The magnetic field is thus saturated at smaller scales. As the turbulent velocity decreases with decreasing scale, also the saturation value for the magnetic field strength decreases correspondingly, thus explaining the maximum at $\sim80$~pc. 

For Burgers-type turbulence, the situation is similar. As the turbulent velocity decreases more rapidly with scale, the magnetic field reaches the saturation level only on somewhat smaller scales of $\sim30$~pc. Towards even smaller scales, it decreases more rapidly, as the turbulent velocity field and thus the saturation field strength decreases more rapidly with decreasing length scale. 

Of course, this estimate is only approximate, as the magnetic field should also be subject to gravitational compression, which reduces the coherence length but amplifies the field strength. In addition, the density will increase during the gravitational collapse, thus increasing the field strength at which the magnetic field saturates. We also expect that the scale on which the turbulence is injected changes during the collapse. If the magnetic field saturates, additional effects like turbulent decay may come into play. To account for these additional physics, we propose a more detailed model in \S~\ref{collapse}.

\section{Magnetic fields during turbulent collapse}\label{collapse}
In this section we derive a model for the evolution of a turbulent gas cloud with initially weak magnetic fields. For this purpose, a model of turbulence and the evolution of the turbulent spectrum during the gravitational collapse is required. In this section, we first develop such a model on the basis of analytical arguments and numerical results \citep{Wise07a,Greif08, Wise08a}.  Based on this model, the growth of the magnetic field in the dynamo phase can be calculated. We also discuss the subsequent evolution in the saturated phase, in which the competition between turbulent decay, gravitational compression and dynamo amplification governs the evolution of the magnetic field.

\subsection{Turbulent collapse}
During the galaxy formation process, turbulence is generated by the release of gravitational energy and the infall of accreted gas on the inner, self-gravitating core. Under such conditions, the injection scale of turbulence is usually comparable to the size of the system under consideration \citep{Klessen09}. Here, we focus on the central density core found in numerical simulations of primordial star formation \citep{Abel02, Bromm03, Yoshida08}. Its extent is comparable to the Jeans length, which can be derived from the critical mass required to make the gas cloud gravitationally unstable:
\begin{equation}
M_J=2M_\odot \left(\frac{c_s}{0.2\ \mathrm{km/s}} \right)^3\left(\frac{n}{10^3\ \mathrm{cm}^{-3}} \right)^{-1/2}.\label{thJeans}
\end{equation}
Here, $c_s$ is the sound speed. This central core is no longer dominated by radial motions, but rather supported  by turbulence generated in accretion shocks. Due to the continuous infall of gas, the turbulence will not decay, but is constantly replenished.  Radial profiles of the turbulent velocity in the first galaxies indicate that there are some random fluctuations, but the order-of-magnitude of the turbulent velocity does not change during the collapse \citep{Wise07a,Greif08, Wise08a}.  For turbulence driven by accretion, one expects that the turbulent velocity is comparable to the infall velocity, and for a roughly isothermal density profile, the free-fall velocity is independent of radius. We will therefore assume that, while the injection scale changes during the collapse, the injected velocity stays the same. On scales smaller than the size of the cloud, we expect that the turbulent velocity scales as a $v\propto l^\beta$. As above, we will explore both Kolmogorov and Burgers-type turbulence. The evolution of the mean density with time is prescribed as in the one-zone models of \citet{Glover09} and \citet{Schleicher09c}, which follow the evolution of primordial chemistry during the collapse phase.

\subsection{Evolution in the dynamo phase}
As discussed in the introduction, there are two mechanisms that can amplify the magnetic field in the collapse phase: The small-scale dynamo, and gravitational compression. In this subsection, we propose a Lagrangian framework that allows to treat these effects simultaneously. We consider an array of fluctuation length scales $l_i$ with a corresponding array of field strength $B_i$, and follow both the evolution of the magnetic field strength $B_i$ and its corresponding length scale $l_i$ over time. For the initial field strength, we adopt a conservative value of $10^{-20}$~G on all scales. 

The initial length scales are chosen with a constant logarithmic spacing, with the largest length scale corresponding to the Jeans length $\lambda_J$, and the smallest length scale corresponding to $10^{-4}\lambda_J$. This range was adopted to make sure that the integral scale always lies well-within the range of scales followed in the code. We have estimated the resistive scale following the framework of \citet{Pinto08a, Pinto08b}, finding that it lies well below this value. We follow the evolution of $1000$ length scales with their corresponding magnetic field strength. For each length scale $l_i$,  we calculate the turbulent velocity $v_i(t)$ as described in the previous subsection.
As the small-scale dynamo amplifies each of these fields on the eddy-turnover timescale, we describe this process by solving the following set of ordinary differential equations (ODEs):
\begin{equation}
\frac{dB_i(t)}{dt}=\frac{B_i(t)}{l_i/v_i}.\label{dynamoODE}
\end{equation}
In addition, gravitational compression amplifies the magnetic field strength and contracts the corresponding length scale. Under spherically symmetric conditions, the magnetic field strength scales as $B\propto\rho^{2/3}$, whereas a more modest increase would be appropriate if the magnetic field could significantly distort the geometry and lead to collapse along one preferred direction. In our model, we consider an initial seed field which is too weak to be dynamically important. Further, even after sufficient amplification, the small-scale dynamo will produce a highly tangled magnetic field, which will also not change the geometry. We therefore increase the magnetic field strength according to this scaling law after each timestep. In addition, the corresponding length scale will be compressed. For collapse under spherically symmetric conditions, we expect it to scale as $l\propto\rho^{1/3}$. During the turbulent collapse, the initial seed field will thus be continuously amplified, while its length scale is  compressed by gravity. 

As the magnetic field within the collapsing cloud is always distorted by the gravitational collapse, the largest coherence length that can be achieved is always smaller than the Jeans length by some factor $f_d$. The precise value of $f_d$ is uncertain, as this problem has not been examined with numerical simulations yet. Throughout this paper, we adopt a fiducial value of $f_d=0.1$. A variation of this value by a factor of  a few will however not change our conclusions significantly. Indeed, even somewhat larger coherence lengths may be possible, but this needs to be explored with numerical simulations. When we determine the integral scale in the subsequent applications, i.e. the scale on which the magnetic field strength is largest, we will thus explicitly ensure that it cannot be larger than $f_d \lambda_J$ . When the saturation field strength is reached, we do no longer solve Eq.~(\ref{dynamoODE}), but switch to the treatment described in the next subsection.

\subsection{Evolution in the saturated phase}
Once the magnetic field $B_i$ on scale $l_i$ is larger than the saturation value given in Eq.~(\ref{maxfield}), it is no longer amplified by the small-scale dynamo. It is however still amplified by gravitational compression according to $\rho^{2/3}$, whereas the saturation field strength only increases as $\rho^{1/2}$. It can thus in principle increase above the saturation level. In this case, it is however subject to turbulent decay. 

In MHD simulations  of decaying turbulence without self-gravity, one typically finds a power-law decay of the total magnetic energy \citep{Smith98, Biskamp99}. For our model, however, we need a prescription for the magnetic field on a given scale rather than the total magnetic energy. As shown e.g. by \citet{Subramanian06}, the power-law behaviour of the total energy results from an exponential decay of the magnetic field on a given scale, which happens on the eddy-timescale. This is because the integral scale increases during the decay, as the energy on smaller scales dissipates more quickly. Due to the growth of the integral scale, the total energy thus decreases as a power-law.

In the presence of gravity, these effects have not been investigated explicitly. However, it seems likely that the picture of purely decaying turbulence will not hold under these circumstances. In particular, it is likely that the integral scale cannot become larger than a fraction of the Jeans length, which we parametrized above as $f_d$. Once saturation is reached, the integral scale within the central core thus continuously decreases over time. 

We therefore do not expect a power-law behaviour as in the case of decaying turbulence without gravity. For the kinetic turbulence, in fact we expect that typical turbulent velocities do not change significantly, although their scale will continuously decrease during the collapse. This is because the turbulence can be continuously replenished by accretion shocks. A magnetic field stronger than the saturation value can however decay. On a given scale, this should occur on the eddy-timescale, similar as for dynamo amplification. We thus describe it as
\begin{equation}
\frac{dB_i(t)}{dt}=-\frac{B_i(t)}{l_i/v_i}.\label{decayODE}
\end{equation}
For a magnetic field strength above the saturation value, its evolution is thus governed by Eq.~(\ref{decayODE}), while it is governed by Eq.~(\ref{dynamoODE}) for weaker fields. As a consequence, the magnetic field thus tends to stay close to the saturation value, which increases as $\rho^{1/2}$.

\section{Application to the first stars and galaxies}\label{application}
The model developed above is now well-suited to study the implications of the small-scale dynamo in the first star-forming systems in the early universe. In this section, we apply it to minihalos that may harbor the first stars, and atomic cooling halos which are often considered to be the first galaxies in the universe. 

\subsection{Evolution in primordial minihalos}
\begin{figure}
\includegraphics[scale=0.5]{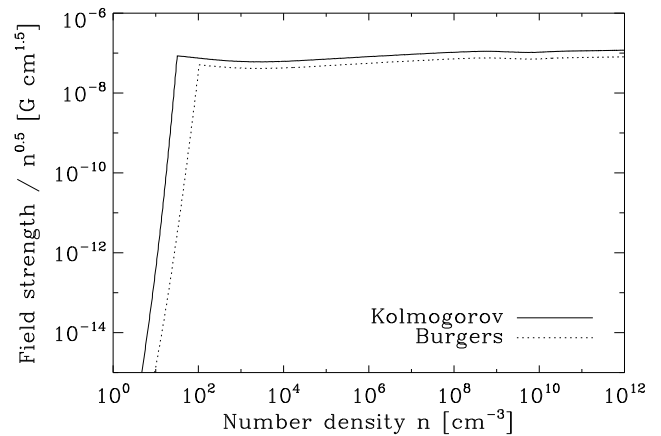}
\caption{The evolution of the magnetic field strength on the integral scale during protostellar collapse in a typical minihalo with subsonic turbulent velocities shown as a function of the average density reached in the central core at a given time. We show cases corresponding to Kolmogorov- and Burgers-type turbulence.}
\label{fig:fieldmini}
\end{figure}

\begin{figure}
\includegraphics[scale=0.5]{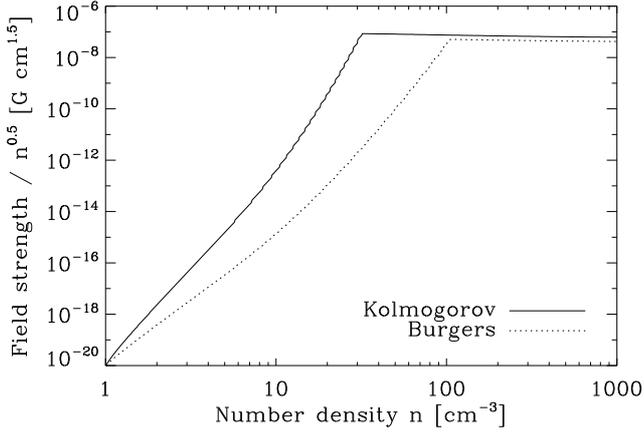}
\caption{Magnetic field amplification in the early collapse phase. We show the same quantities as in Fig.~\ref{fig:fieldmini}, with particular focus on the initial phase.}
\label{fig:fieldminizoom}
\end{figure}

\begin{figure}
\includegraphics[scale=0.5]{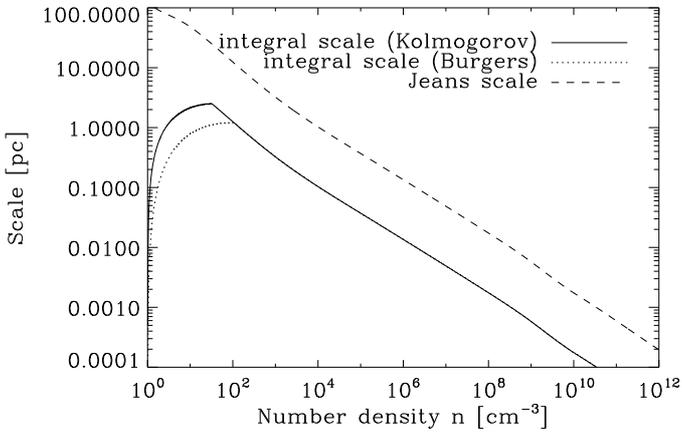}
\caption{The evolution of the integral scale of the magnetic field as a function of the average density reached in the central core at a given time during protostellar collapse in a typical minihalo with subsonic turbulent velocities. We show cases corresponding to Kolmogorov- and Burgers-type turbulence and compare with the Jeans scale.}
\label{fig:integralmini}
\end{figure}

\begin{figure}
\includegraphics[scale=0.5]{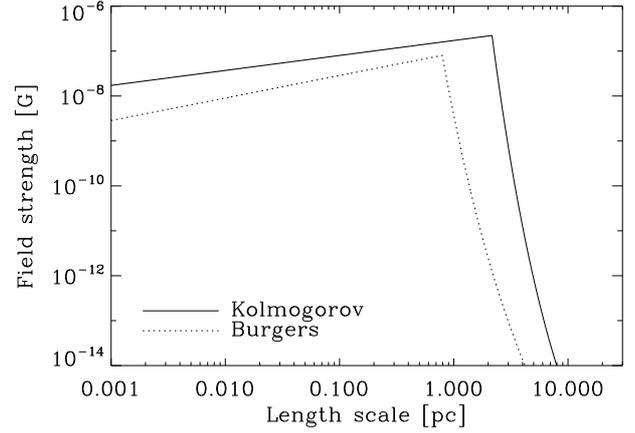}
\caption{The magnetic field as a function of scale when an average density of $10$~cm$^{-3}$ is reached in the central core, in a typical minihalo, shown for Kolmogorov- and Burgers-type turbulence.}
\label{fig:minispec}
\end{figure}

\begin{figure}
\includegraphics[scale=0.5]{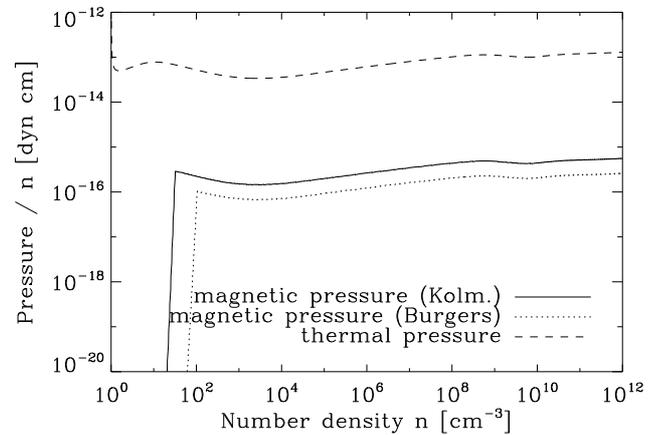}
\caption{Thermal and magnetic pressure as a function of the average density reached in the central core at a given time during protostellar collapse in a typical minihalo with subsonic turbulent velocities, for Kolmogorov- and Burgers-type turbulence. In both cases, thermal pressure dominates over the magnetic one.}
\label{fig:pressuremini}
\end{figure}

The first stars are suggested to form in primordial minihalos with $\sim10^5-10^6\ M_\odot$ between $z\sim30$ and $z\sim15$ \citep{Abel02, Bromm04, Glover05, Yoshida08}. These halos have initial temperatures of a few $1000$~K, and an initial ionisation degree of $\sim2\times10^{-4}$, corresponding to the relic electron fraction left after recombination \citep{Seager99, Seager00}. The main cooling mechanism is thus line emission from molecular hydrogen, which forms primarily due to the $H^-$ channel. 

The turbulent properties in the first star-forming minihalos have not yet been studied to a satisfactory degree and thus provide a relevant uncertainty in this analysis. The presence of turbulence is however evident from the inhomogeneities in the central cloud cores found in a number of simulations, and the sub-keplerian angular momentum profiles \citep{Abel02, Bromm03, Yoshida08}.  \citet{Tan04} mentions the presence of shock-velocities comparable to the sound speed. 

In our turbulent collapse model, we will thus assume that a turbulent velocity equal to the sound speed is injected on the Jeans scale, leading to sub-sonic turbulence on the integrale scale of the magnetic field. We consider both Kolmogorov and Burgers-type turbulence. 

With the model obtained in \S~\ref{collapse}, we follow the evolution of the magnetic field spectrum, and plot the evolution of the magnetic field strength on the integral scale, as well as the integral scale as a function of density, in Figs.~\ref{fig:fieldmini} and \ref{fig:integralmini}. In the initial dynamo phase, the magnetic field strength increases rapidly by several orders of magnitude, as the eddy-turnover time on sub-pc scales is much smaller than the free-fall timescale. We show the evolution during this dynamo-phase in more detail in Fig.~\ref{fig:fieldminizoom}. The spectrum of the magnetic field during this early phase, at a density of $10$~cm$^{-3}$, is shown in Fig.~\ref{fig:minispec}. As one can see, the integral scale is initially small, but increases up to $\sim3$~pc at a density of $\sim10^2$~cm$^{-3}$.~At that point, the magnetic field is in the saturated phase, where it increases roughly with $n^{0.5}$, while the integral scale decreases as the Jeans scale due to gravitational compression. We find the same qualitative behaviour for the Kolmogorov- and the Burgers-type turbulence, indicating that the precise scaling of turbulent velocity with length scale is of minor importance. In both cases, we find a rapid build-up phase and a subsequent saturation phase. For Burgers-type turbulence, the build-up is just slightly delayed, and the field saturates on a slightly lower level, as the typical velocity on the largest possible integral scale, $f_d \lambda_J$, is somewhat decreased. 

In Fig.~\ref{fig:pressuremini}, we show the evolution of magnetic pressure for both cases. We find that the thermal pressure clearly dominates over the magnetic pressure in both cases. Magnetic fields created by the small-scale dynamo may thus not change the protostellar collapse phase significantly. However, they are strong enough to change fragmentation behaviour and binary formation in the disk phase, as discussed in \S~\ref{conclusions}. Magnetic effects therefore need to be considered for a correct assessment of the primordial initial mass function (IMF).

\subsection{Evolution in atomic cooling halos}
\begin{figure}
\includegraphics[scale=0.5]{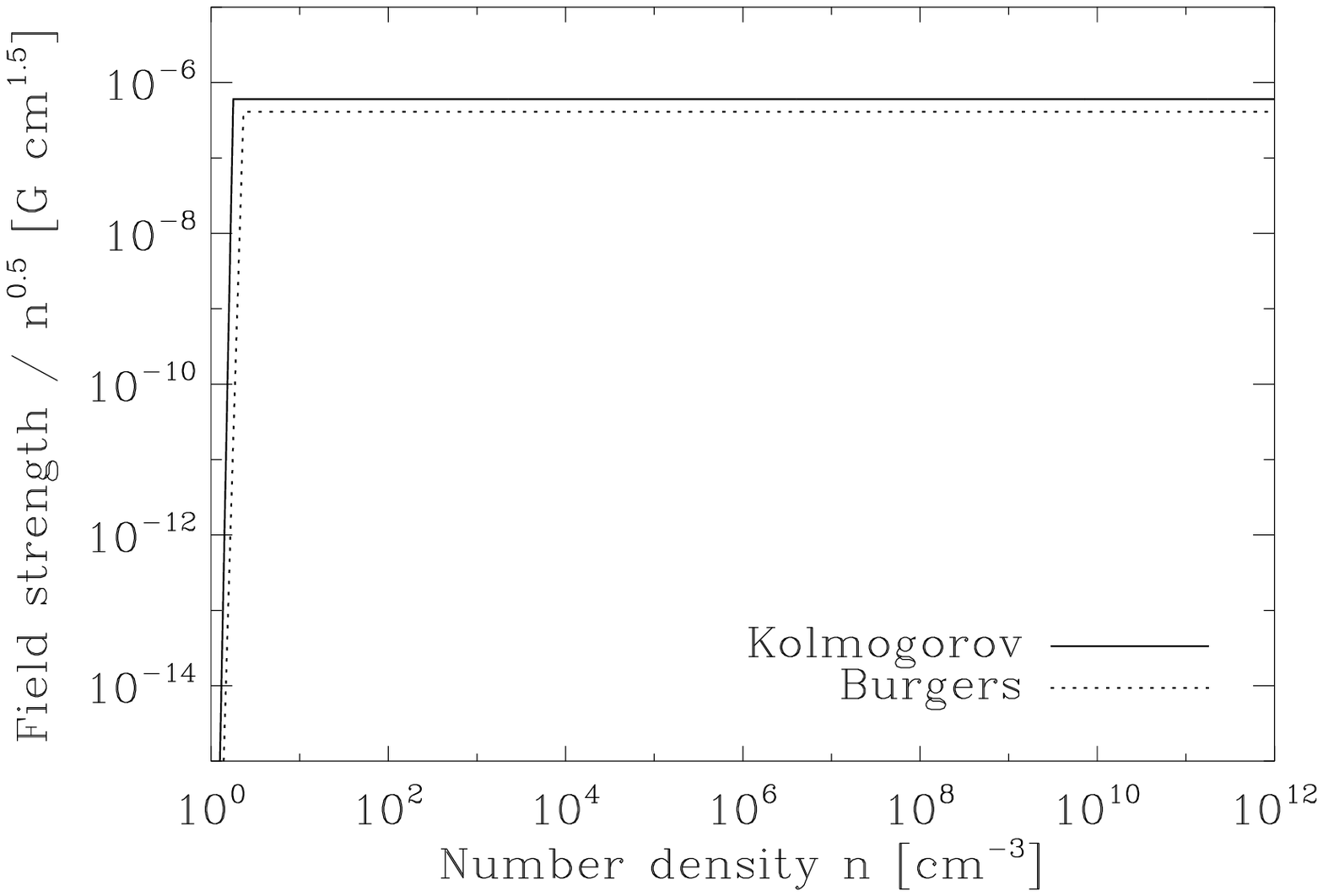}
\caption{The evolution of the magnetic field strength on the integral scale during protostellar collapse in an atomic cooling halo with turbulent velocities of $\sim20$~km/s on the injection scale shown as a function of the average density reached in the central core at a given time. We show cases corresponding to Kolmogorov- and Burgers-type turbulence.}
\label{fig:fieldatomic}
\end{figure}

\begin{figure}
\includegraphics[scale=0.5]{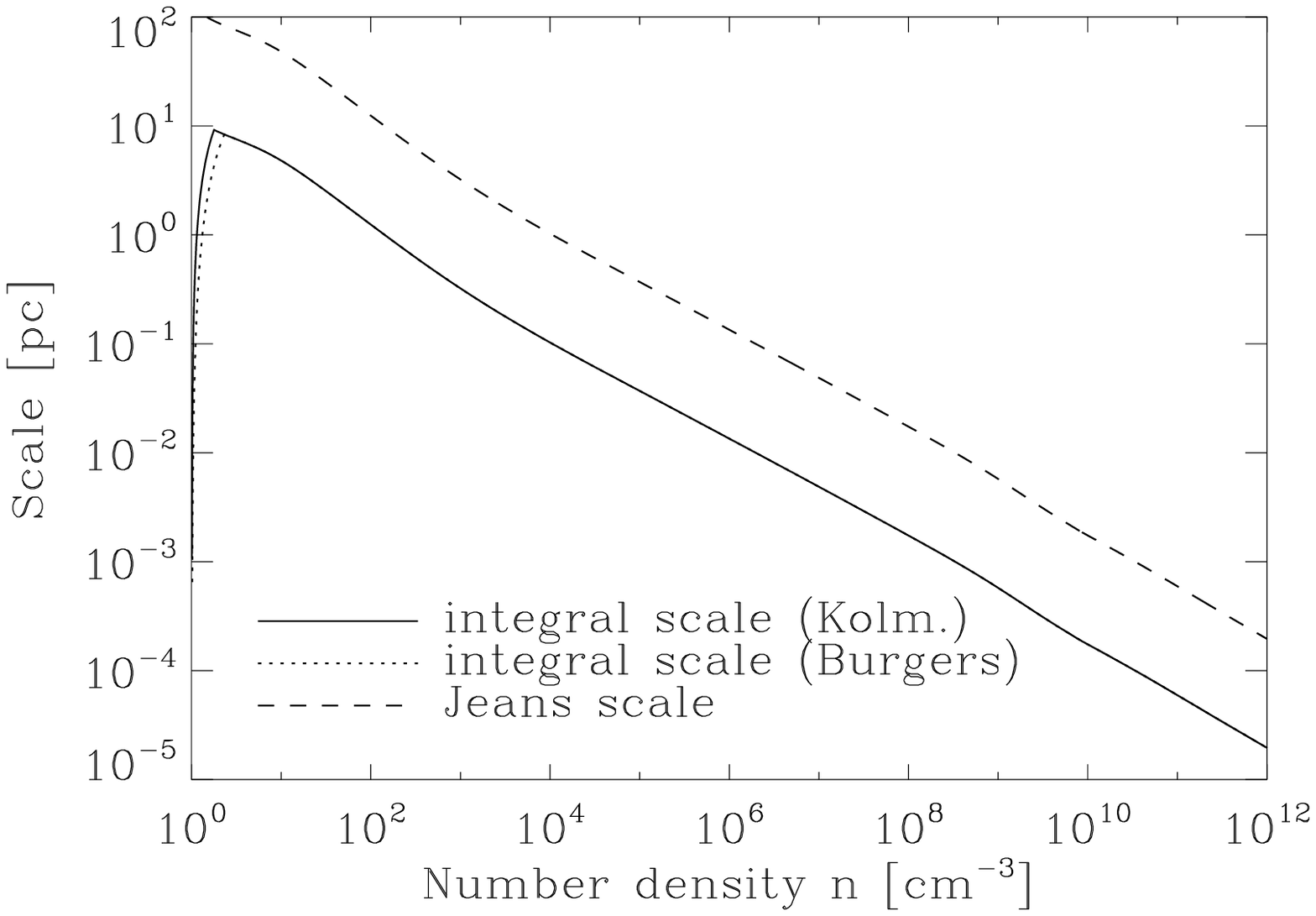}
\caption{The evolution of the integral scale of the magnetic field as a function of  the average density reached in the central core at a given time during protostellar collapse in an atomic cooling halo with turbulent velocities of $\sim20$~km/s on the injection scale. We show cases corresponding to Kolmogorov- and Burgers-type turbulence and compare with the Jeans scale.}
\label{fig:integralatomic}
\end{figure}

\begin{figure}
\includegraphics[scale=0.5]{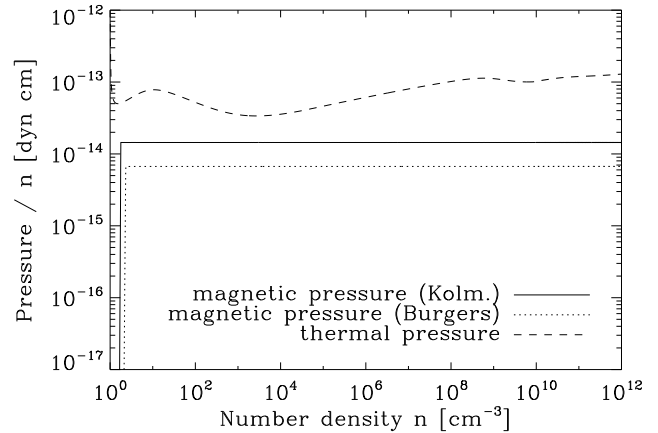}
\caption{Thermal and magnetic pressure as a function of the average density reached in the central core at a given time during protostellar collapse in aan atomic cooling halo with turbulent velocities of $\sim20$~km/s on the injection scale, for Kolmogorov- and Burgers-type turbulence.}
\label{fig:pressureatomic}
\end{figure}
Atomic cooling halos are defined as systems with virial temperatures of at least $10^4$~K and have masses of $M > 5 \times 10^{7} [(1 + z) / 10]^{-3/2} \: {\rm M_{\odot}}$. Graviational infall can thus shock-heat the gas to $\sim10^4$~K, leading to an increased ionisation degree up to $10^{-2}$. Under such conditions, atomic hydrogen becomes an important cooling agent. In addition, the presence of an additional cold accretion mode has been found in numerical simulations \citep{Greif08}, which results form accretion in dense filaments with enhanced fractions of molecular hydrogen. The gas can thus cool efficiently while it is accreted, thus staying at temperatures of a few hundred K. Due to the large column densities, Lyman $\alpha$ photons may be efficiently trapped in these systems \citep{Spaans06}. The presence of a photodissociating background may suppress the $H_2$ abundance and increase the gas temperatures \citep{Omukai01}. Even the combination of these effects is however not able to prevent the gas temperature from decreasing during the collapse \citep{Schleicher10b}. 

As shown by \citet{Greif08} and \citet{Wise07a,Wise08a}, the gas in these systems is subject to supersonic turbulence. They find supersonic turbulence with Mach numbers up to $10$, corresponding to turbulent velocities of $\sim20$~km/s. We will  adopt this as a fiducial value for the turbulent velocity on the injection scale and again consider the consequences of Kolmogorov and Burgers-type turbulence. 

As shown in Fig.~\ref{fig:fieldatomic}, the magnetic field strength increases even more rapidly in this case, due to the supersonic turbulent velocities in atomic cooling halos. The magnetic field saturates at densities still lower than $10$~cm$^{-3}$ and evolves then with $n^{0.5}$, like the saturation scale. As shown in Fig.~\ref{fig:integralatomic}, the integral scale also increases more rapidly, yielding values of $\sim10$~pc at a density of $\sim3$~cm$^{-3}$. Towards higher densities, its evolution is dictated by gravitational compression. It thus scales as the thermal Jeans length. As for minihalos, the difference between a Kolmogorov and a Burgers-type spectrum is of minor importance, as in both cases the saturation value is rapidly reached. 

In Fig.~\ref{fig:pressureatomic}, we compare magnetic and thermal pressure in atomic cooling halos. Due to the supersonic turbulent velocities, the saturation field strength is highly increased, and thus also the magnetic pressure. The difference between the thermal and the magnetic field strength varies between $0.5$ and one order of magnitude. In the presence of additional coolants due to metal-enrichments, this difference will decrease further.
As discussed below in more detail, the magnetic field will be highly inhomogeneous in a three-dimensional configuration and may thus be dynamically important locally. The magnetic field may thus change the evolution in the collapse phase and also after the formation of a disk. Potential consequences are discussed in more detail below.

\section{Discussion and conclusions}\label{conclusions}\
We demonstrated in this paper that magnetic fields are generated rapidly by the small-scale dynamo both in minihalos which form the first stars, as well as in atomic cooling halos which may harbor the first galaxies. In this section, we summarize the main results and discuss the main consequences and open questions which cannot be resolved in this semi-analytic framework. 

\subsection{Formation of the first stars}
The formation of the first stars was often examined by hydrodynamical simulations that neglected potential effects from magnetic fields \citep[e.g.][]{Abel02, Bromm04, Yoshida06}. Magnetic fields have however been considered to be important in the protostellar disk in presence of an efficient dynamo \citep{Tan04, Silk06}. 
{The impact of a uniformly imposed magnetic field ($B_0$) in the primordial collapse has been 
studied by \citet{Machida06, Machida08} using direct numerical MHD simulations. In these simulations,
a range of $B_{0} = 10^{-6} - 10^{-9}$G was used for different values of the angular velocity
of rotation. A protostellar jet within a radius of about 0.02 AU was found to be launched for an
initial $B_{0}\geq 10^{-9}$~G at $n=10^3$ cm$^{-3}$.} In this paper, we showed that the small-scale dynamo leads to a magnetic field strength much larger than the critical value of $10^{-9}$~G~$(n/10^3$~cm$^{-3})$ derived by \citet{Machida06} for the formation of jets and outflows. However, as these magnetic fields are more tangled than those of \citet{Machida06}, their results cannot be directly applied to ours, and additional numerical studies concerning tangled magnetic fields are required. The average Alfv'en velocity $v_A=B/\sqrt{\rho}$ is typically smaller than the sound speed, though it may dominate locally because of fluctuations in the magnetic field strength.

In addition, our study clarifies that the magnetic field strength in the protostellar disk is much higher than previously anticipated. The conditions for the MRI, as formulated by \citet{Tan04} and \citet{Silk06}, are thus fulfilled. This leads to the presence of turbulence in the protostellar disk, which may drive a large-scale $\alpha\omega$ dynamo in the presence of some kinetic helicity, making the field stronger and more coherent. But also the MRI itself may further amplify the magnetic field \citep{Balbus91}.

This has important consequences for the fragmentation behaviour of the disk. Detailed numerical studies of the collapse of magnetised  molecular cloud cores in the  context of present-day star formation \citep[e.g.][]{HennebelleT08, HennebelleF08, HennebelleC09, Mellon09} indicate that even modest field strengths can suppress binary formation and strongly favour the formation of single stars. Jets and magnetic tower flows are very effective in transporting away angular momentum and thus change structure and dynamics of the protostellar accretion disk. { On the other hand, numerical simulations exploring the interaction of turbulence generated by the MRI with gravitational instabilities indicate the excitation of additional modes and an effective reduction of the accretion rate, as well as the broadening of spiral arms by the MRI turbulence \citep{Fromang04}. Dedicated numerical studies exploring the combination of such effects in a primordial accretion disk will thus be required to understand the full impact on the stellar masses.
}


\subsection{Formation of the first galaxies}
The formation of the first galaxies is currently subject to much larger uncertainties than the formation of the first stars. This is because the initial conditions are not completely clear and the amount of metal enrichtment is not fully understood. The presence of supersonic turbulence has however been convincingly demonstrated by \citet{Greif08} and \citet{Wise08a} with cosmological simulations encorporating hydrodynamics and primordial chemistry. Additional physics like supernova feedback may just enhance the amount of turbulence found there. The small-scale dynamo is found to be extremely efficient under these conditions and may magnetise the material during the collapse, with the magnetic pressure only half an order of magnitude below the thermal pressure on average. Locally, the magnetic field may even dominate in some places, as it is expected that highly inhomogeneous fields are generated from the small-scale dynamo. Indeed, as discussed by \citet{Subramanian99} and \citet{Brandenburg05}, the magnetic field is highly inhomogeneous, reaching equipartition in about $10\%$ of the volume. The saturation field strength we adopted above results from a spatial average over the different local values. We thus expect fluctuations of the field strength by at least a factor of $10$ \citep{Wang09, Dubois09}. Due to the increase of the integral scale, the magnetic field becomes more coherent in these systems, making a stronger case for the putative presence of jets and outflows. 

As in the case of minihalos, the formation of a disk may lead to the presence of an $\alpha\omega$ dynamo that makes the magnetic field more coherent on disk scales. It may similarly play a role by making angular-momentum transport more efficient and thus reducing the amount of fragmentation and suppressing binary formation. As discussed above, the magnetic pressure may locally dominate over the thermal pressure. In this case, the magnetic Jeans mass sets the critical scale for fragmentation. For small-scale turbulent fields, it is defined in analogy to the thermal Jeans mass as
\begin{equation}
M_{J,B}=2M_\odot \left(\frac{v_A}{0.2\ \mathrm{km/s}} \right)^3\left(\frac{n}{10^3\ \mathrm{cm}^{-3}} \right)^{-1/2}\propto \frac{B^3}{{\rho^2}}.\label{thJeans}
\end{equation}
Here, the Alfv'en speed $v_A=B/\sqrt{4\pi\rho}$ replaces the sound speed $c_s$, as magnetic pressure support propagates with the Alfv'en speed. With
\begin{equation}
v_A=2.0\ \mathrm{km/s} \left( \frac{B}{10^{-5}~\mathrm{G}} \right) \left( \frac{n}{10^2\ \mathrm{cm}^{-3}}  \right)^{-0.5},
\end{equation}
the Alfv'en speed in the saturation phase is thus larger than or comparable to the speed of sound.

The presence of such fields thus provides additional stability during the formation of intermediate-mass black holes, which are often considered to form in such systems \citep[e.g.][]{Eisenstein95, Koushiappas04, Begelman06, Spaans06, Shang09,Schleicher10b}. Thus, even if fragmentation cannot be totally avoided in hydrodynamical simulations, the presence of magnetic fields may still give rise to larger seed masses. The detailed consequences however need to be assessed with numberical simulations. Additional open questions concern the further evolution of the magnetic field on larger scales and the build-up of galactic-scale fields, as discussed by \citet{Arshakian09}. Their model for the magnetic-field evolution in galaxies yields a number of predictions which can be tested with future radio facilities such as the SKA\footnote{http://www.skatelescope.org/}, which can thus constrain the formation mechanisms of the small- and large-scales magnetic fields in new born and young galaxies. In this respect, cosmological simulations that include an approximate treatment for the mean-field induction equation, as performed by \citet{Dubois09}, will be very important.

An additional issue that needs to be addressed is the role of magnetic helicity. Magnetic helicity is a conserved quantity and affects magnetic field generation and decay. In the presence of helical fields, the small-scale dynamo also creates correlations on larger scales \citep{Subramanian98, Subramanian99}. The decay law for helical fields was derived by \citet{Hatori84}. It is independent from the large scale part of the spectrum (i.e. scales above the integral scale) and is generally less efficient due to helicity conservation. During such decay, magnetic power is shifted from small to large scales via an inverse cascade, thus increasing the typical coherence length \citep{Christensson01}. Our paper was conservative in the sense that we assumed magnetic fields with zero helicity, yielding a lower limit on the integral scale. In the presence of helicity, magnetic fields may be coherent on larger scales, making it more straightforward to drive large-scale jets and outflows. To assess this issue, the turbulent properties of the first galaxies need to be analyzed and understood in further detail.

\subsection{Further discussion}
Based on the estimates performed for this paper, it seems likely that the small-scale dynamo will be very efficient during the formation of the first stars and galaxies. The epoch of first star formation may thus also be the epoch where the first strong magnetic fields formed in the universe. This may be important for our understanding of primordial star formation. We further speculate that this mechanism may not only apply to the very first galaxies, but that the formation of any gravitationally bound structures lead to a sufficient amount of accretion-driven turbulence to amplify magnetic fields. We plan to investigate this proposition further with numerical simulations.

\section*{Acknowledgments}
We thank Axel Brandenburg, Daniele Galli, Marita Krause, Ralph Pudritz, Dmitry Sokoloff and Kandu Subramanian for stimulating discussions on the topic { and the anonymous referee for helpful remarks on our manuscript}. The research leading to these results has received funding from the European Community's Seventh Framework Programme (/FP7/2007-2013/) under grant agreement No 229517. Robi Banerjee is funded by the Emmy-Noether grant (DFG) BA 3607/1. {RSK thanks the German Science Foundation (DFG) for support via the Emmy Noether grant KL 1358/1.} DRGS and RSK also acknowledge subsidies from the DFG SFB 439 {\em Galaxies in the Early Universe}. DRGS, RSK, SS and TGA thank for funding via {the Priority Programme 1177 {\em"Witnesses of Cosmic History:  Formation and evolution of black holes, galaxies and their environment"} of the German Science Foundation}. In addition, RSK  thanks for subsidies from the German {\em Bundesministerium f\"{u}r  
Bildung und Forschung} via the ASTRONET project STAR FORMAT (grant  
05A09VHA) and from the {\em Landesstiftung Baden-W{\"u}rttemberg} via  
their program International Collaboration II. RSK also thanks the KIPAC at Stanford University and the Department of Astronomy and Astrophysics at the University of California at Santa Cruz for their warm hospitality during a sabbatical stay in spring 2010. { KIPAC is sponsored in part by the U.S. Department of Energy contract no. DE-AC-02-76SF00515.}


\end{document}